\documentclass[12pt]{article} 
     \usepackage{amsmath}
  \usepackage{graphicx}
  \newlength{\abstractwidth}
  \newcommand{\be}{\begin{equation}}
  \newcommand{\bea}{\begin{eqnarray}}
  \newcommand{\eea}{\end{eqnarray}}
  \newcommand{\beq}{\begin{equation}}
  \newcommand{\ee}{\end{equation}}
  \newcommand{\eeq}{\end{equation}}

  \newcommand{\half}{{1\over 2}}
  
\def\la{\label}

\def\32{{3 \over 2 } }

  \def\ba{\begin{eqnarray}}
  \def\ea{\end{eqnarray}}

 \def\simleq{\; \raise0.3ex\hbox{$<$\kern-0.75em
      \raise-1.1ex\hbox{$\sim$}}\; }
 \def\simgeq{\; \raise0.3ex\hbox{$>$\kern-0.75em
      \raise-1.1ex\hbox{$\sim$}}\; }

\def\nref#1{(\ref{#1})}

\begin{document}

\begin{titlepage}
  \bigskip

  \bigskip\bigskip

  \bigskip

\begin{center}
{\Large \bf { 
A  model with cosmological Bell inequalities 
 }}
 \bigskip
{\Large \bf { }}
    \bigskip
\bigskip
\end{center}

  \begin{center}

\bigskip
\bigskip

 {Juan Maldacena }

\bigskip 
\bigskip

 { \it  School of Natural Sciences, Institute for Advanced Study, \\
 Princeton, NJ, USA\\ }

\bigskip
\bigskip

  \end{center}

 \bigskip\bigskip
  \begin{abstract}
We discuss the possibility of devising cosmological observables which violate
Bell's inequalities. Such observables could be used to argue that cosmic scale features were 
produced by quantum mechanical effects in the very early universe. 
As a proof of principle, we 
 propose a somewhat elaborate  inflationary model where a Bell inequality violating observable
can be constructed.

 \medskip
  \noindent
  \end{abstract}
\bigskip \bigskip \bigskip
 
\bigskip 
\begin{center} 
{\it Dedicated to Andy Strominger on the occasion of his $60^{th}$ birthday.}  
\end{center}

  \end{titlepage}



\section{Introduction}

 According to the theory of  inflation,  primordial density fluctuations have a quantum mechanical origin
\cite{Mukhanov:1981xt,Hawking:1982cz,Guth:1982ec,Starobinsky:1982ee,Bardeen:1983qw}. 
An important problem is to find compelling evidence for their quantum nature. 
In other words,   one would like to rule out alternative scenarios where the fluctuations originated 
through classical statistical mechanics during an inflationary phase.  
This can happen in models where there is a  form of friction  converting the inflaton energy into 
 other forms of energy that then produce the fluctuations as in \cite{Berera:1995wh,Berera:1995ie,LopezNacir:2011kk,Senatore:2011sp}. 
In such theories, the   correlations between the fluctuations are classical in origin. 
One can compare detailed predictions for higher point correlation functions, or even put bounds on the 
two and three point functions, see e.g.  \cite{Mirbabayi:2014jqa}.

 Formally, the wavefunction of the universe produced by inflation is highly entangled. 
Therefore one   expects
 that it should be possible to perform a Bell inequality violating  experiment \cite{Bell:1964kc}.  
Such an experiment    would conclusively demonstrate the quantum origin of the fluctuations. 
 Here we will ask  the  conceptual question of  whether, and in what sense, could we 
 ever perform such a cosmological  Bell experiment.  
 A Bell experiment involves making measurements at two distant locations, call them  Alice's and Bob's location.  
 At each of these locations one should be able to measure two non-commuting operators.  
 In cosmology, 
 we can make observations on two spatially separated cosmic  patches that  have been
 causally disconnected since the  time of reheating.
  However, it is not possible to measure
 two non-commuting operators for the following reason.
  The standard observables involve measuring the values of the cosmological curvature fluctuations (or
adiabatic density fluctuations),  
 $\zeta(x)$. However,  it is not possible to 
 measure its  conjugate momentum,  $\pi_\zeta$. \footnote{
 Bell inequalities were discussed previously in  \cite{Campo:2005sv}, but with the assumption that one 
 can indeed measure the momentum $\pi_\zeta $.}
 
 At this stage one could conclude that it is impossible to perform a Bell type measurement in cosmology. 
 But this would be premature.  
  First notice that    {\it any} observation we make consists of  commuting observables, once
 we consider only the final  decohered observables \cite{Hartle:1992as}. 
 In order to run the Bell experiment one repeats the experiment many times, 
 interpreting each run of the experiment as occurring on the same quantum state, and putting the boundary between classical 
 and quantum just after the measurements. 
 We can view  cosmology in a similar way.  We can view separate   patches of the sky  as running different cosmological 
 experiments on the same underlying quantum state. 
 We can further  divide  these  patches into a pair of smaller subregions, which were causaly disconnected at 
 some earlier time.  By suitably observing properties of these subregions one could construct an observable 
 subject to a Bell inequality.  There is a Bell inequality if one assumes that 
  the probability distribution that we observe today was 
 generated by an inflationary process leading to a relation between scale and time. Namely short distance features were created 
 after the long distance features.    In this case, we can 
 translate the standard causality constraints in the 
 approximately de-Sitter space into constraints on the spatial structure of the wavefunction. 
 
 We have been unable to find an observable of this kind using the simplest inflationary theory consisting of the metric
 plus  a single scalar field. However, we will present a more baroque inflationary scenario where 
 one can prove the quantum origin of some fluctuations.  This scenario was solely designed to make a Bell inequality 
 violating experiment possible and it seems unlikely that Nature would choose it. 
 We think it is nevertheless valuable  to have a fairly concrete model where one can clearly understand the 
 various issues involved. Hopefully, a clever reader (or non-reader) 
 will find a  Bell inequality violating  observable in a more realistic model. 
 
 This paper is organized as follows. First, in section two, 
 we  review the standard discussion of Bell inequality experiments. In section three 
we review a few features of inflation and discuss the conceptual set up for a cosmological Bell
experiment.  In section four we 
  present a baroque inflationary model where a Bell inequality violating  experiment is possible. 
 We conclude with a discussion. 

 \section{ Review of Bell inequality experiments } 
\la{ReviewBell}
 
\begin{figure}
\begin{center}
 \includegraphics[scale = 0.27]{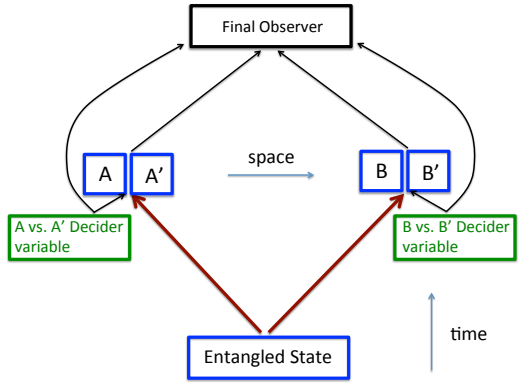}

\caption{ Set up for a Bell inequality violating experiment. An entangled state is produced in the past and its two parts are transmitted to  Alice and 
Bob who perform measurements on $A$ or $A'$ or $B$ or $B'$. The choice of experiment ($A$ vs. $A'$)
 is determined by a local variable, which we can call Alice's ``free will'' or
decider variable.   The results of the experiments and the values of the decider variables 
are classically transmitted to a central observer who computes the statistical averages. All classical communications 
have been denoted here by black lines.  }\label{fig1}
\end{center}
\end{figure}

 In order to set up an experiment with a Bell inequality it is necessary to have the following elements
\cite{Bell:1964kc}. See figure \ref{fig1}.
 \begin{itemize}
 \item
 Two separate spatial locations where measurements are performed. Call them   Alice's location and Bob's location. 
 \item
 An entangled quantum state, with components at these two locations. 
 \item
 At each location we should be able to perform two possible measurements that are described by two 
 non-commuting operators. Call them $A$ and $A'$ for Alice's location and $B$, $B'$ for Bob's location,
with $[A,A'] \not = 0 $ and $[B,B']\not =0$. 
 \item
 Alice  should have the ``free will'' to select randomly between the $A$ and $A'$. The same holds for Bob for his choice of  $B$ and $B'$
 These choices are made locally and are uncorrelated with each other. These choices are made by physics
outside the quantum system under consideration. In practice this is done by looking at local random variables that are assumed to be independent of the 
quantum system in question.\footnote{  There is a Bell inequality violating measurement involving    $B$  $\bar B$  oscillations
 \cite{Go:2003tx} where the validity of this  assumption has been called into question \cite{Bertlmann:2004cr,Ichikawa:2008eq}.}
 \item
 We should have a quantum measurement of these operators with definite answers. 
 \item
 We classically transmit the results of these measurements
 to a central location where we correlate the results. 
 \end{itemize}
  
  Let us review the simplest and most discussed example. Here the entangled state corresponds to a pair of spins, one at each location. 
  The operators correspond to measuring the spin along various axes and have eigenvalues $\pm 1$. In other words, 
  we have that $A = \vec n . \vec \sigma = n^i \sigma^i$, with $\sigma^i $ the Pauli matrices. And $A' = \vec n' . \vec \sigma$. We have similar expressions for
  $B$ and $B'$ acting on the second spin. 
  
 In this situation is it useful to consider the quantity introduced in \cite{CHSH} 
 \be \la{Cdef}
\langle  C \rangle =  \langle A B \rangle   + \langle A B' \rangle   + \langle A' B \rangle  - \langle  A' B' \rangle 
 \ee
 This is a particular linear combination of expectation values for different choices of operators or detector settings. 
 
 In a local classical hidden variable theory one can prove the Bell inequality
  $ |\langle C \rangle | \leq 2$ as follows. 
  For each value of the hidden variables we have a well defined answer for each of the two possible measurements at each side. Namely, a 
  unique value  for $A$ and also for $A'$, similarly for $B$ and $B'$.   
  Furthermore, causality implies that 
  the answer for $B$ and $B'$ does not depend on whether we measure $A$ or $A'$. Therefore, for each value of the hidden variable
   we can have either $B = B'$ or $B = - B'$. 
  And in each of the two cases either the first two terms in \nref{Cdef} cancel or the last  two terms cancel.
   Therefore the maximum value of $|C|$ is two.  
  
  In quantum mechanics, the expectation value of $C$ can be bigger. In fact, in quantum mechanics we can view \nref{Cdef} as the expectation value of the 
  quantum operator $C = A B + A B' + A' B - A' B'$. It is easy to check that its square is 
\be
  C^2   =   4 -   [ A,A'][B,B']  
\ee
where we used that the square of each of the measured operators is one, $A^2 =1$, $A'^2 =1$, etc. 
Now the commutator term can make $C^2$ larger than four. Only when this commutator is non-zero can we have $|\langle C\rangle |$ larger than two, violating
the Bell inequality.  
Notice also that $|[A,A']| \leq 2.$\footnote{This inequality is saturated for Pauli matrices. For example,
 consider  $A= \sigma_x$ and $A' = \sigma_y$, which leads to
$[\sigma_x , \sigma_y] = 2 i \sigma_z$.}. Therefore it is easy to see that  $\langle C^2 \rangle \leq 8$ or $|\langle C \rangle |\leq 2\sqrt{2}$ \cite{Cirelson}. 
Choosing 
\be \la{choices} A = \sigma_x ~,~~~A' = \sigma_y ~,~~~~B =  \sin \theta \sigma_x + \cos \theta \sigma_y~,
~~~~~~~~B' = \cos  \theta \sigma_x  -\sin \theta \sigma_y
\ee
 we can check 
that on a spin singlet state we get $C = - 2 \sqrt{2}  \cos ( \theta - { \pi \over 4} ) $.  For $\theta = \pi/4$ we get the maximal violation which has the extra 
$\sqrt{2}$ factor.

There has also been discussion of Bell inequalities for harmonic oscillator degrees of freedom. For example,
 \cite{BaWo} considers squeezed states and 
 defines operators that are translation conjugates of $(-1)^n$  where  $n$ is the
occupation number operator. 
These  operators  are not   easy to measure in the cosmological context. They certainly cannot 
be measured  after reheating   \cite{Campo:2005sv}, due to the impossibility of measuring $\pi_\zeta$. 
 For a more realistic inflationary scenario, it will be necessary
to consider entangled states and measurements of the harmonic oscillator degrees of freedom that describe the scalar or tensor fluctuations.
 However, for our baroque model this discussion in terms of spins will be  enough.

 \section{Set up for a cosmological Bell experiment }
 
 \subsection{Review of inflation } 
 
\begin{figure}
\begin{center}
\includegraphics[scale = 0.35]{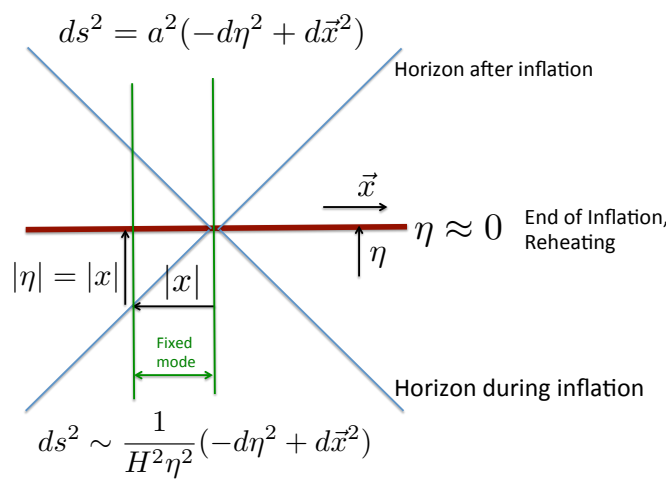}
\caption{  Sketch of the evolution of the universe. The vertical direction is conformal time, $\eta$, and the horizontal is space. 
 We have an early period of inflation ending at $\eta\sim 0$ followed by an  ordinary radiation/matter dominated  universe. 
  A comoving distance $x$ crosses the horizon during inflation at time $\eta \sim - |x|$. So scales 
correspond to time. The vertical green line follows a given wavelength mode as it crosses the apparent horizons given by the diagonal lines. 
 }\label{UniverseEvolution}
\end{center}
\end{figure}

 We can write the metric of a uniform, spatially flat,  FLRW space as 
 \be \label{metran}
 ds^2 = - dt^2 + a(t)^2 d\vec x^{\, 2} =  a^2(\eta)[ - d\eta^2 +   d \vec x^{\, 2} ]  
 \ee
 where $t$ is proper  time and $\eta$ is conformal time. The $\vec x$ coordinates are called ``comoving coordinates". 
Conformal time is particularly useful to display causal relations. 
During the inflationary period the scale factor grows exponentially,  ${\dot a \over a} \sim H(t) $, with $H(t)$ slowly varying.  
 The variation of $H(t)$ is due to the slow evolution of a 
 scalar field, which in the classical  approximation is a function of time only $\phi = \phi_0(t)$.
 Inflation ends when $\eta \sim 0$, see figure \ref{UniverseEvolution}.  For a review, see e.g. \cite{Linde:2005ht}. 
 
  Quantum mechanics  produces spatially dependent fluctuations in the values of the scalar field 
  \cite{Mukhanov:1981xt,Hawking:1982cz,Guth:1982ec,Starobinsky:1982ee,Bardeen:1983qw}. These 
  give rise to adiabatic  curvature fluctuations in the late universe. 
  In the leading approximation,  we can independently follow the evolution of each Fourier mode, $\phi_{\vec k}(\eta)$. Each of these Fourier modes
  behaves as a  harmonic oscillator with a time dependent mass and fixed frequency.
   Each mode corresponds to a wave whose wavelength is fixed in the $\vec x$ coordinates of \nref{metran}.
  Their physical wavelength is very small at early times and very large towards the end of inflation. When $k |\eta| \sim 1$ the fluctuations are created and the 
  value of $\phi_k$, or more properly that of $\zeta_k =- { H \over \dot \phi_0 M_{pl} } \phi_k $, 
   is fixed until the mode reenters the horizon during the Big Bang phase, see figure \ref{UniverseEvolution}. The non-constant part of $\zeta$ decays exponentially in 
   proper time after we exit the horizon. Furthermore, the amplitude of the second independent solution decays as $ (\eta k)^3$ after horizon exit. This is also the
   order of magnitude of the commutator of $k^3 [\zeta_{\vec k} , \dot \zeta_{-\vec k}] \propto i (\eta k)^3 $. This goes as $e^{ - 3 N_k}$ where $N_k$ is the number of
   e-folds that remain  from the time the mode exited the horizon till the end of inflation\footnote{ In a general FLRW background
    this commutator goes as $[ \zeta , \dot \zeta ] \propto a^{-3}$ so 
   that it becomes even smaller after the end of inflation and subsequent horizon reentry, since the universe continues expanding.}.  
   For cosmological size modes this is a number bigger than about $N_k > 30-40$. 
   Therefore, if we wanted to make a measurement of the momentum we would need a  precision greater than $10^{-90}$ which,
   even for a theorist,  looks impossible. Moreover, the measurement of any cosmological observable is limited by cosmic variance which goes as $1/\sqrt{N_{mod}}$ where
   $N_{mod}$ is the number of modes we observe. Even if we observe all modes up to the size of a galaxy, this gives us an ultimate precision of about $10^{-10}$. 
  
  In summary, inflation gives us a probability distribution for $\zeta(x) $ at the time of reheating of the form $\rho[\zeta(x)] = \left| \Psi[\zeta(x)] \right|^2$. 
  Since we cannot measure  the decaying mode,  we can view the state of the universe at the reheating surface as 
  characterized by the classical probability distribution  $\rho[\zeta(x)]$. 
    If we consider two well separated points $x_A$ and $x_B$ then the operators $\zeta(x_A)$ and $\zeta(x_B)$ commute with each other. Therefore
   it is impossible to obtain a Bell inequality out of these operators.\footnote{See
 \cite{Nambu:2011ae,deAlwis:2015ioa} for a 
related discussion. }

        If more than one field is involved, then
  we can also have isocurvature fluctuations, but the conclusion is the same. 
 The probability distribution is classical and  has the form  $\rho[\zeta(x),\theta(x)] = |\Psi(\zeta(x), \theta(x) )|^2$, where $\theta(x)$ is the second field. 
   
  Therefore, if we view   Alice and Bob as doing experiments after the end of inflation, then we will not be able to set up a Bell inequality for primordial perturbations. This is true if we make the realistic assumption
that we cannot measure the conjugate momentum. 
  This is disappointing! But fortunately this is  not the end of the story.

    \begin{figure}
\begin{center}
\includegraphics[scale = 0.27]{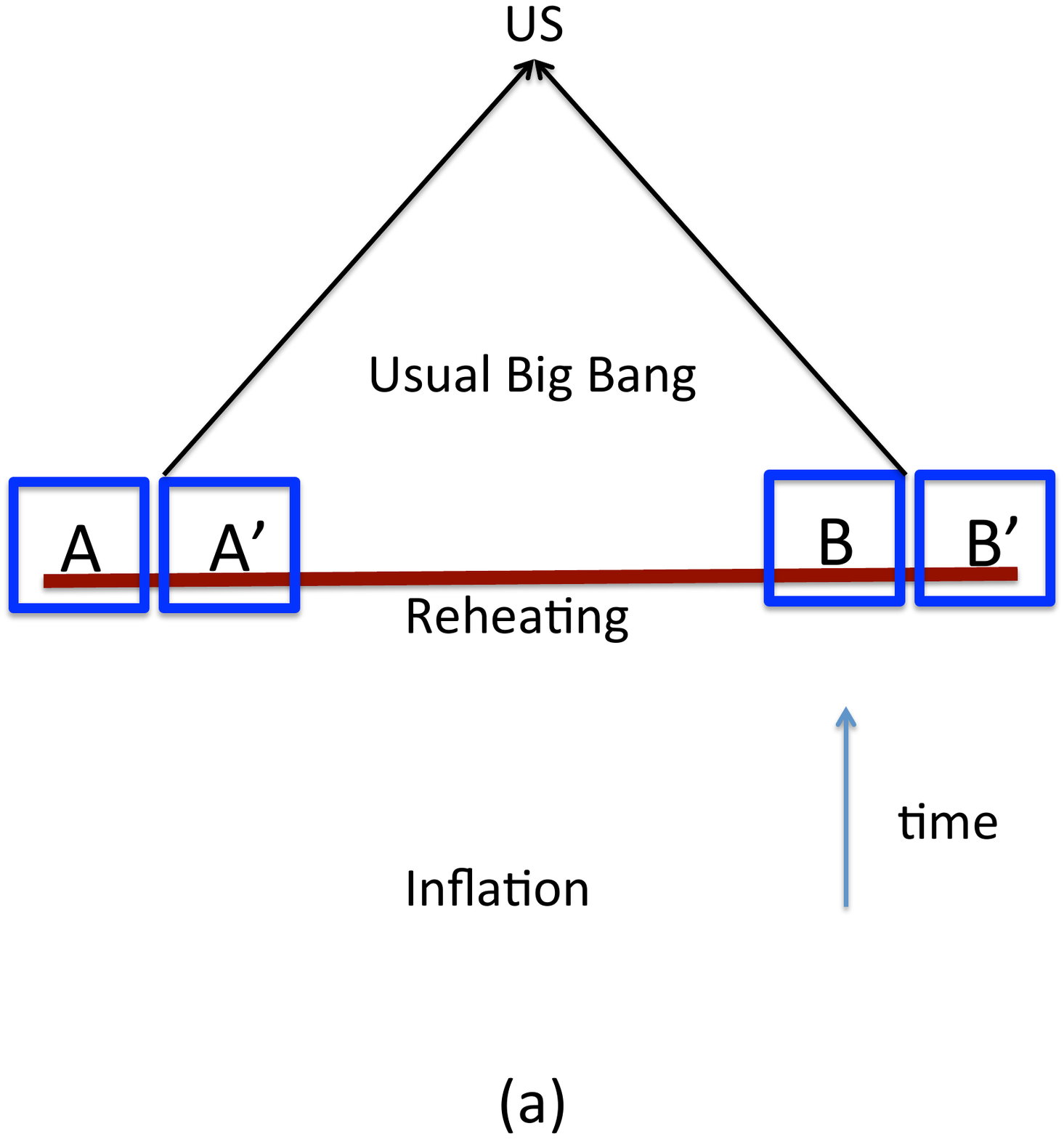}~~~~~~~\includegraphics[scale = 0.27]{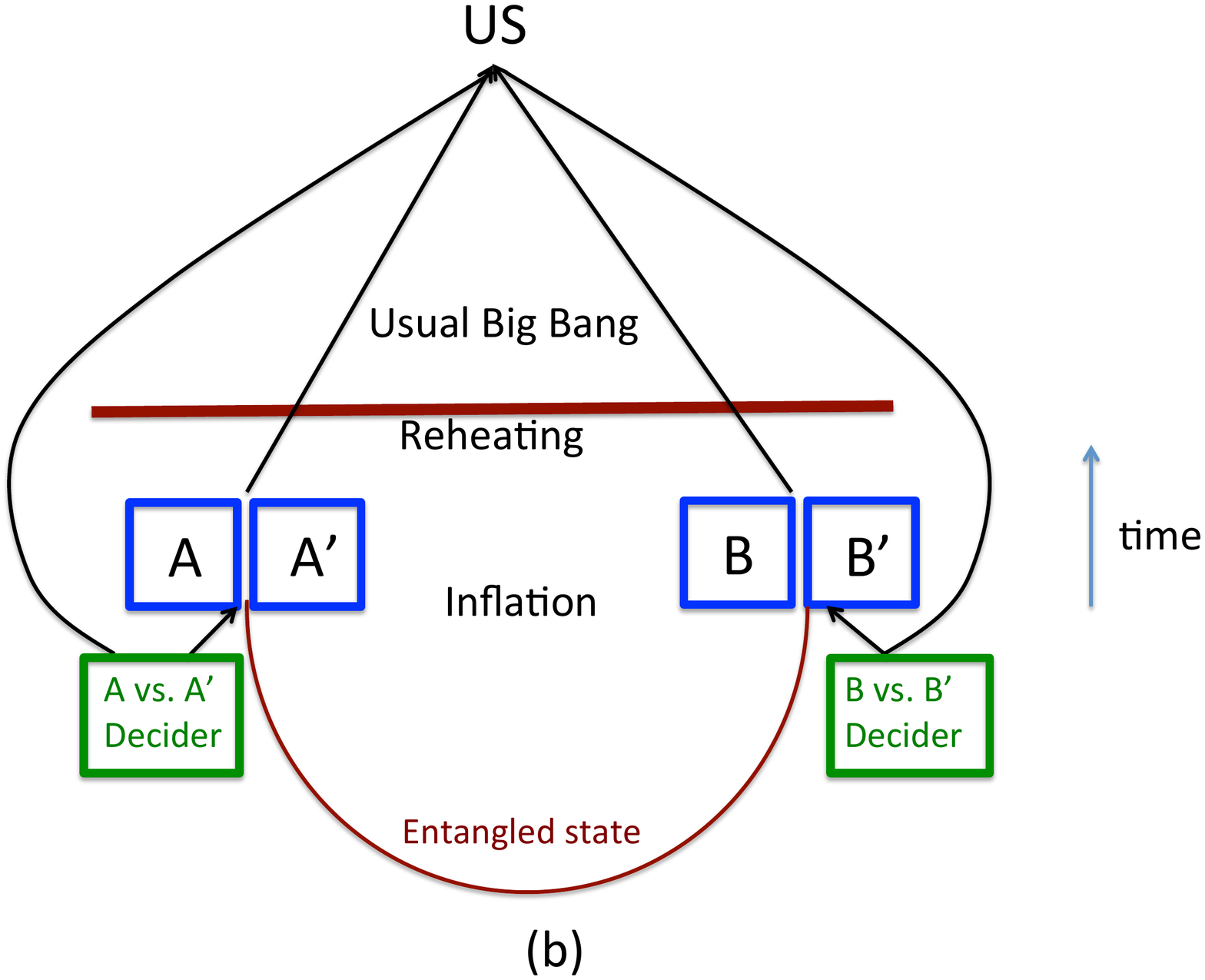}
\caption{ (a) An unsuccessful set up for a cosmological Bell inequality. There are no non-commuting operators that we can   measure after the end of inflation.  
(b)  A diagram of a more successful set up where the whole process occurs during inflation. 
 We generate an entangled state. Some time later we  generate the variables that will decide whether we make an $A$ or $A'$ measurement and similarly for
$B$ and $B'$. These decider variables as well as the result of the measurement should remain as classical 
variables for the rest of the evolution and be visible to us.   }\label{SetUp}
\end{center}
\end{figure}

  \subsection{ Setting up a cosmological Bell experiment} 
  
  Implicit in any discussion of the Bell inequality is the assumption of when we make the quantum to classical transition. In other words, when
  the measurement occurs. For example, even in the standard setup of figure \ref{fig1} it is important that the quantum to classical transitions 
  happen in such a way that
  we can view all  black lines as classical. 
  Similarly, in cosmology, we can 
    avoid the previous conclusion if we  imagine an Alice and a Bob who did their experiments during inflation and 
  ``wrote'' their results on the classical distribution $\rho[ \zeta(x) , a(x)]$. In other words, this classical distribution is viewed as the classical message which is 
  transmitting to us the result of experiments which happened during inflation, see figure \ref{SetUp}(b).

   Now, what are suitable Alices and Bobs?. Of course, they do not need to be actual people.
    More important than Alice and Bob  are the measurements that they do. 
    A measurement is a particular unitary evolution of the combined system plus measuring apparatus, whose 
  state can be viewed  as classical. 
  We need  to produce all the elements of the Bell inequality discussion out of fluctuations. The initial entangled state would be a quantum fluctuation, 
   the measurement apparatus would be another quantum fluctuation that has already become classical. It should have shorter wavelength than the one corresponding
   to the entangled state. This shorter wavelength fluctuation should act both as the decider variable as well as measuring apparatus. The measurement should be
   some process which depends on the quantum state of  one of the pieces  of the entangled sate.  The result of the measurement should be transmitted to us. 
   Therefore the measurement should be some process which produces a large effect on the fluctuations so that we can see it today. The state of the shorter wavelength
   fluctuations that acted as ``decider'' variables should also be preserved and transmitted to us. We know   one mechanism for transmitting this information. 
   Namely, through  the inflationary evolution of massless (or nearly massless) scalar fields where small fluctuations are amplified and stretched to cosmic scales. 
   In figure \ref{SetUp}(b)  we sketch the type of setup that we have in mind. 
   
     Unfortunately, we have not been able to produce 
   a suitable observable  using the simplest single scalar field model. One difficulty is the following. 
   We mentioned above that  the fluctuations become classical as they exit the horizon. The fluctuations which will serve as the detector and decider variables 
   are necessarily of shorter wavelength that the ones in the entangled state. This is in order to ensure that the values of the decider variables are determined 
   locally, independently for Alice and Bob. Unfortunately this also means that the entangled state we are attempting to measure is actually more classical than the
   measuring device, which is the opposite of what we want.

 \section{A baroque model that leads to a Bell inequality measurement} 
 
Instead of giving up, we will imagine that we have a more complicated model of inflation
 where we can indeed set up a Bell inequality. 
   Simply as a matter of principle, we would like to ask whether there is an inflationary model that is Bell-friendly. Namely,   one that spontaneously creates,
    from the vacuum, 
     all the necessary elements for the Bell-inequality experiment,     performs the measurement and records the results for the post-inflationary observer. 
 One can imagine several ways of doing this, 
but we will concentrate on one particular example in order to display a concrete model. 
 
  The model  builds the various elements in the Bell experiment  as follows 
  \begin{itemize} 
  \item
  The entangled state consists of a pair of massive particles that carry an isospin degree of freedom. The isospin is entangled in the singlet state.
 These particles are very massive at the beginning of inflation, then they get lighter at a specific time and then they become heavier again. Therefore
  they are created at a specific time during inflation, the time when they become lighter. 
  \item
  The decider variables or detector settings correspond to an axion field.
 The axion field  has a variable decay constant, $f_a$. 
Again, this is large for most of the time, but it becomes smaller, comparable to $H$ for a few efolds. 
This means that the dimensionless axion angular variable has larger fluctuations at a particular 
scale.   
  The axion field survives beyond the end of inflation and  
  produces isocurvature fluctuations. These isocurvature fluctuations retain the information of the detector settings for the post-inflationary observer. 
  \item
  The measurement occurs as follows.  We postulate  an isospin dependent contribution to the mass of the particles. This contribution  arises from a term
  that becomes important a few Hubble times after that pair is created. This coupling depends on the axion value. The measurement consists in an interaction 
  between the mass of this particle and the inflaton. 
  \item
  The result of the measurement is preserved for the post-inflationary observer as follows. 
  The massive particles   classically modifies  the evolution of the inflaton so as to 
  produce a discernible fluctuation, or hot spot,  larger than the quantum fluctuations. 
  These hot spots are centered where the massive particles are located. Their amplitude depends on 
 the projection of the  isospin of the massive particle along an axis whose 
direction is axion dependent.   
 \end{itemize}

 The final conclusion is that this baroque universe has produced a very particular pattern of curvature and isocurvature fluctuations. The pattern of curvature
 fluctuations is mainly the usual  almost scale invariant one with additional hot spots  where there is a significant deviation.   
 These spots  come in well separated 
 pairs. Each separated pair constitutes a particular instance of an entangled pair together with a measurement. The measurement operation depends on
 the value of the axion field, which  can be read off from the isocurvature fluctuations. 
  
 In order to have a more clear example we now discuss these elements in more detail.

   \subsection{ Creation of well separated pairs of massive particles}
   \la{CreationPairs}

    \begin{figure}
\begin{center}
\includegraphics[scale = 0.8]{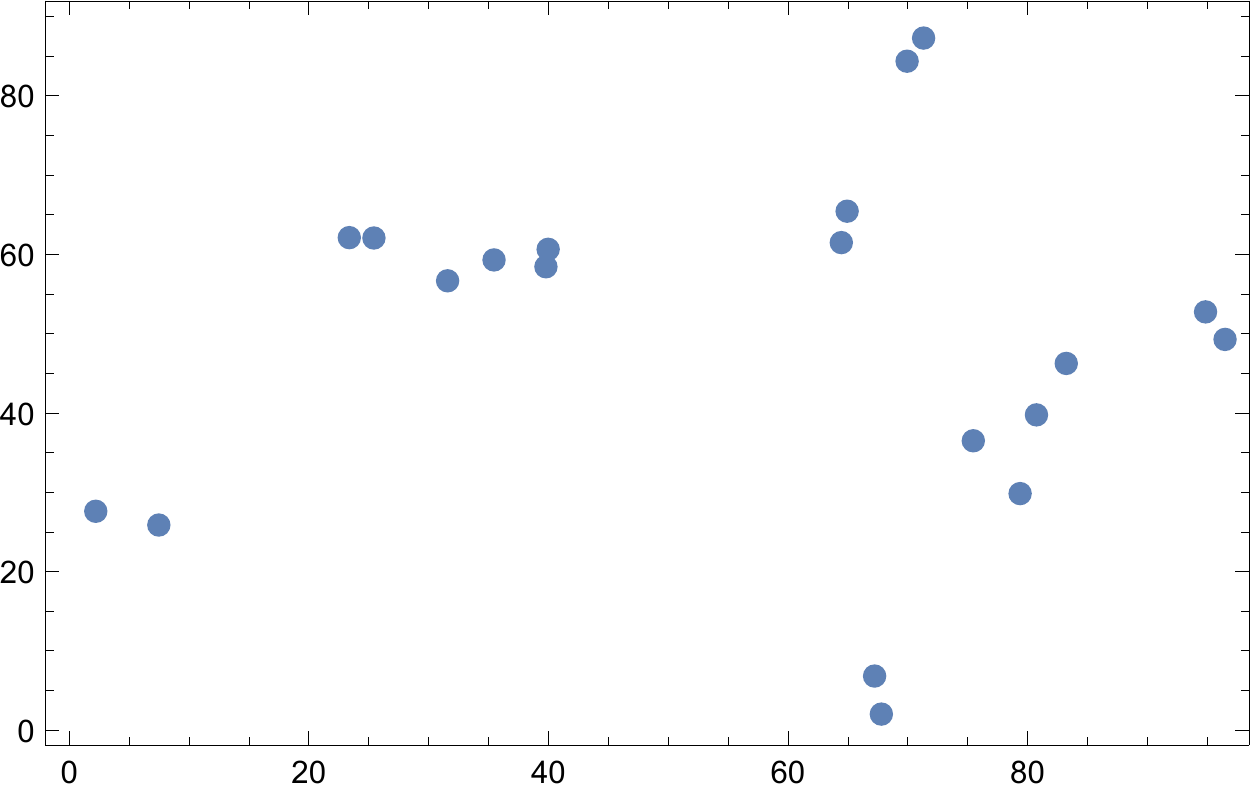}
\caption{ One instance of the particle creation process with the statistics that follows from a time 
dependent mass during inflation. The center of mass position of the pair has a uniform random 
distribution.
 The distribution of distances  between the 
pairs peaks at a distance that is set by the time during inflation where the particles became less massive. 
The axes are comoving coordinates. We see that it is 
reasonably easy to recognize the members of a particular pair. }\label{masspairs}
\end{center}
\end{figure}

    We imagine massive particles whose mass depends on the inflaton $\phi$, $m(\phi)$. The particles are generically very massive, 
    $m\gg H$. But we also imagine that there is a particular value of the inflaton, $\phi_0$, where the particles become relatively light
    $  H \sim  m $, but somewhat bigger than $H$ so that we do not produce too many of them. As the inflaton evolves,  there is a particular time
    when it passes through $\phi_0$. At this time  particle pairs are created. We are interested in a 
     situation where the pair creation is rare enough that particles are
    well separated, but strong enough that we produce lots of pairs in the observable universe. It is useful to think of the background metric as approximated by 
    \be 
    ds^2 = H^{-2} {  - d \eta^2 + d \vec x^{ \, 2 } \over \eta^2 } 
    \ee
    The classical time dependence of  the inflaton leads to a time dependent mass $m(\eta)$. 
    The equation of motion for the massive field is 
    \be \la{massparteq}
      h'' - { 2 \over \eta } h' + ( k^2 + { m^2(\eta) \over \eta^2 H^2  } ) h =0 
      \ee
      We imagine a situation where the WKB approximation is approximately valid for all times. If the WKB approximation were exactly valid, there would be 
      no particle creation. We can
      consider a small amount of particle creation which is characterized by a Bogoliubov coefficient $\beta$ which tells us the mixing between the two 
      WKB solutions. Here $\beta(k)$ is small in the region $ k|\eta_0| < 1$ and it is very small for larger values. This leads to a probability distribution for the
      relative comoving distance $x$ between the two pairs which peaks at $x \sim |\eta_0|$. 
      Of course,  the probability distribution for the center of mass of the pair is completely uniform. A simulation of such a distribution is given in figure \ref{masspairs}. 
      We give a bit more details in appendix A. 
           The important feature here is that the typical  distance $x$ (in comoving coordinates) is of the order of the time at which the pair is created.

      \subsection{ Axion with time dependent $f_a$} 
      
\begin{figure}
\begin{center}
\includegraphics[scale = 0.8]{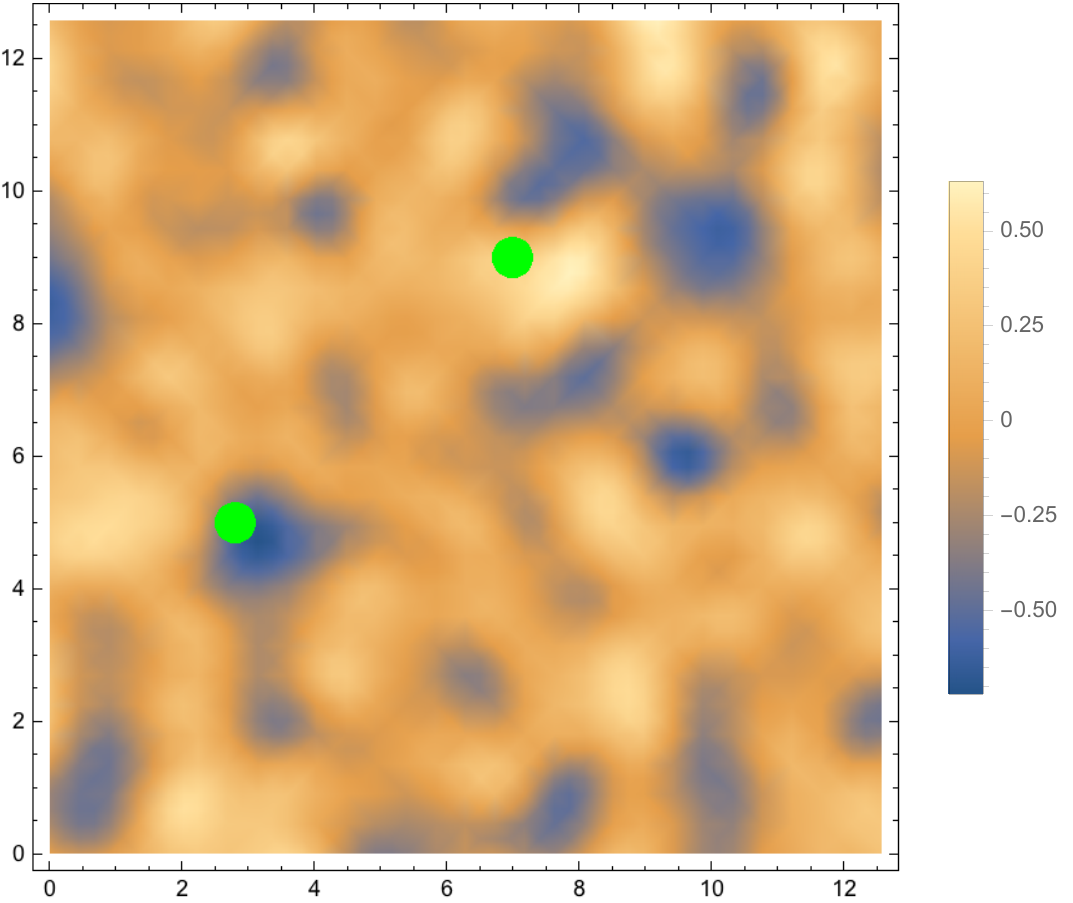}
\caption{ Here we see a profile of the generated axion field. It has features on characteristic scales which 
are small compared  to the separation between the massive particles, so that each massive 
particle sees a different
value of the axion when the ``isospin'' of each particle is measured. The green dots are two members of a particular pair of created particles. This figure is zoomed relative to figure \ref{masspairs}.  }\label{axionandmasses}
\end{center}
\end{figure}

      In this subsection we imagine an axion field with an action 
      \be \la{actaxion}
       S = \int f_a^2 (\nabla \theta)^2  = \int d\eta d^3 x {  f_a^2(\eta) \over H^2 } {[  (\partial_\eta \theta)^2 - (\partial_i \theta)^2 ]\over \eta^2 } 
      \ee
      where $\theta$ is a periodic field, $\theta = \theta + 2 \pi$.\footnote{ We are using the word ``axion'' to describe a periodic field, but this field does not
      have to be the QCD axion.} 
      We assume that the axion ``decay constant''  $f_a$ depends on the inflaton $\phi$. Since $\phi$ is time dependent, then 
      $f_a$ becomes time dependent. We  assume that $f_a$ starts out large, $f_a \gg H$, 
       and becomes smaller, but larger than  $H$ at $\phi_1$ and that then it rises and 
      becomes  large again. 
      The net effect of this is that the fluctuations of the angular variable $\theta $ are larger  at distances corresponding to time $\eta_1$, 
      $x \sim |\eta_1|$,  and  then they become much smaller at shorter distances. Simulated axion fluctuations with these properties can be found in figure \ref{axionandmasses}. 
      It is conceptually cleaner to imagine that 
       $\phi_1 $ comes at a slightly later time than  $\phi_0$ discussed in section \ref{CreationPairs}, but no great harm is done if they  happen together.  
        But it is important that $f_a$ rises for later times.   
         In figure \ref{axionandmasses} we show both the axion field and the created particle pairs, 
zooming on a particular pair.
 We see that each member of the pair can be in regions with different values of
         the axion field. The evolution of the axion was fine-tuned to generate rather different values of $\theta$ at the locations of each member of the pair. 
         The increase of $f_a$ has allowed us to suppress quantum fluctuations at shorter distances so that for the next step we have a 
         well defined value for  the axion field at the location of each particle. 

Also, since the axion has fluctuations at distances shorter than the separation between the massive pairs, 
we can view the fluctuations around each pair as being produced locally. In other words, since the 
axion becomes the decider variable, we want to ensure that it is chosen locally around each massive particle,
in a way that independent from what happens around the other massive particle of the pair. 
 This happens in this 
model if we view the generation of the fluctuations as occurring when the modes cross the horizon.\footnote{
Of course, since the whole region under consideration started out in a   subhorizon region, one can call this
independence 
assumption into question. Of course, this assumption can also be called into question, for the same reason, 
in present day Bell experiments (even \cite{Gallicchio:2013iva}) since the whole observable universe was initially contained in a small 
Hubble region. In other words, we are applying for inflation the same kind of assumptions we are used to 
applying for present day experiments. } 
         
         We imagine that the axion field survives beyond  the end of inflation  and that there is a small potential $ V \sim \Lambda^4 \cos \theta$ which  gives it a mass and
         causes it to oscillate at late times. This contributes as a dark matter component.
The axion fluctuations give rise to   isocurvature fluctuations in this dark matter component. 
 In our universe we do not see such fluctuations in the
         overall dark matter density, but this could be a subdominant contribution of the dark matter. 
         Here the point is that, in principle, by observing the size of these isocurvature fluctuations we can determine the initial amplitude of the field $\theta$ in the
         corresponding regions of the universe.  Now in order not to produce domain walls we want that $f_a$ remains always smaller than $H$ so that the axion field
         is always around the same minimum of the potential.  Indeed,
 in figure \nref{axionandmasses}
we see that fluctuations are smaller than $\pi$.

     \subsection{ Seeing the massive particles after the end of inflation} 
      \la{seeingmassive}
      
      The particle pairs that we discussed above will be diluted by the expansion of the universe and one can wonder how they will ever be observable. 
      First we will discuss a mechanism that makes them observable. Later  
  we will modify the mechanism  to  include the 
      isospin degree of freedom. 
      
\begin{figure}
\begin{center}
\includegraphics[scale = 0.6]{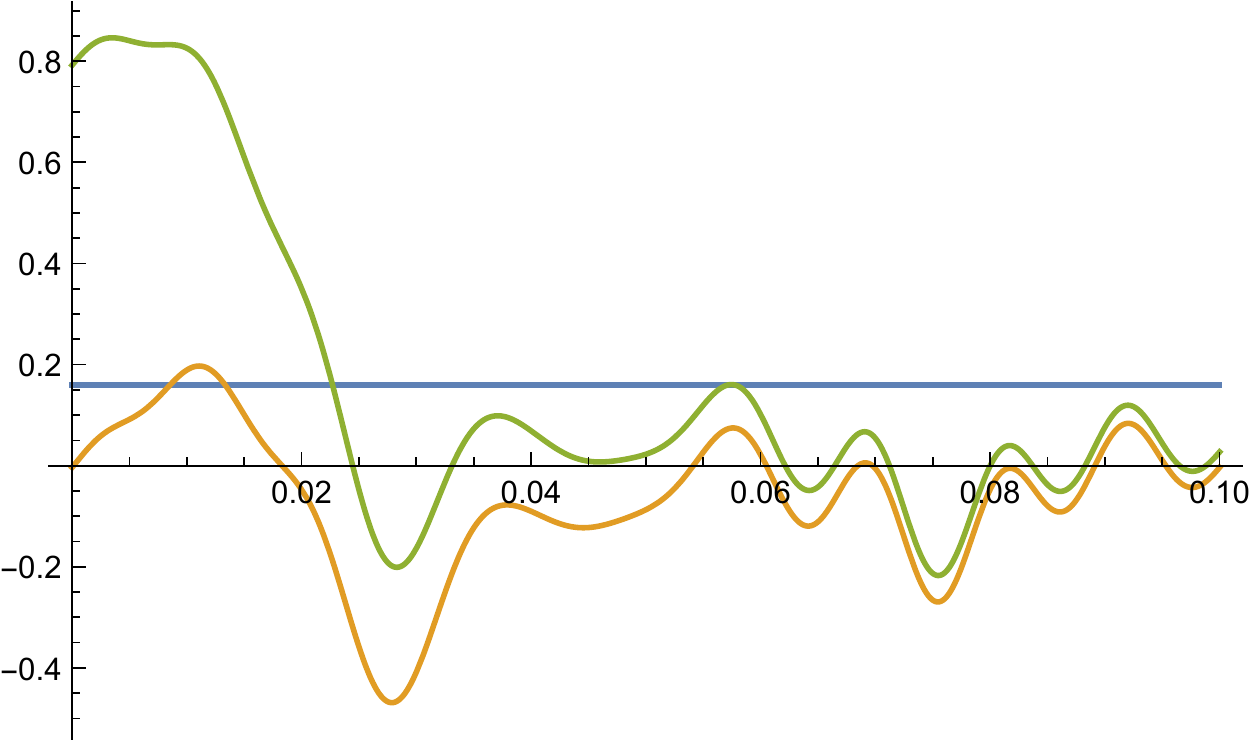}~~~~~~\includegraphics[scale = 0.8]{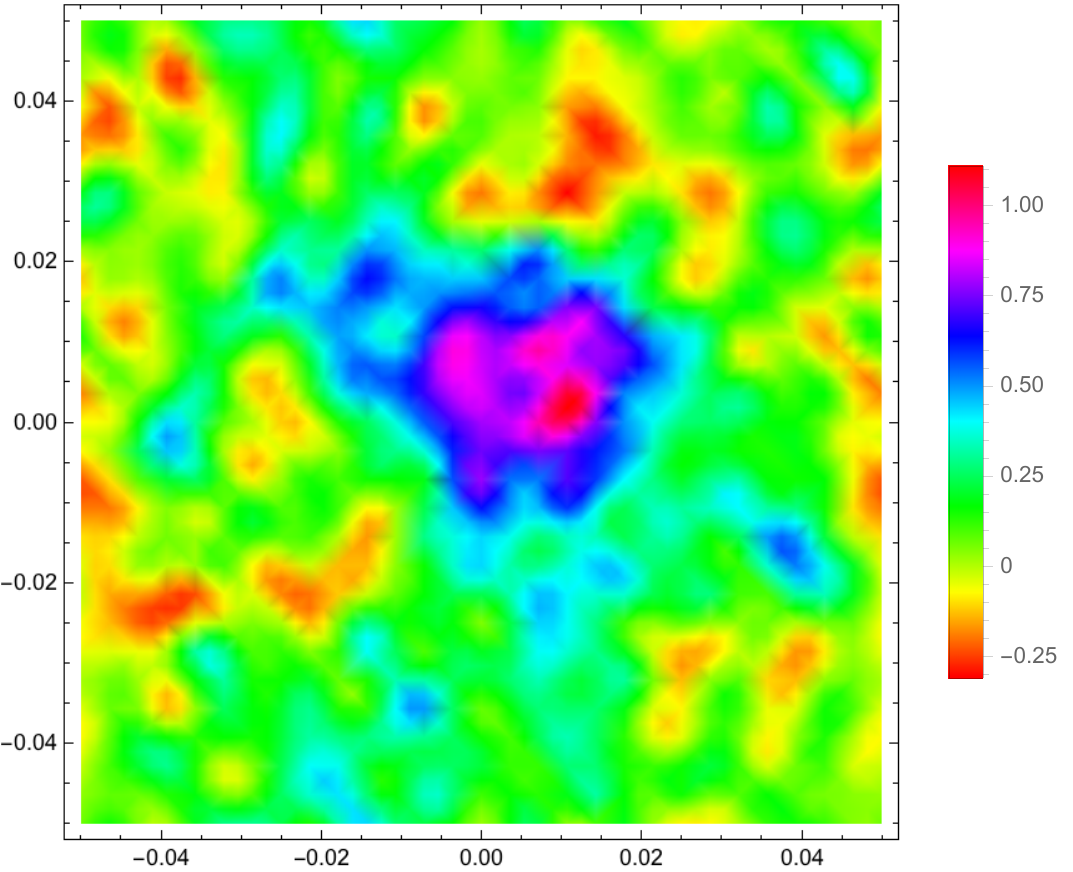}
(a) ~~~~~~~~~~~~~~~~~~~~~~~~~~~~~~~~~~~~~~~~~~~~~~~~~~~~(b)
\caption{ For these figures we assumed a particular time dependence of the mass where $\langle \zeta_{part} \rangle$ 
 becomes a factor of about five above the background value of the fluctuations (see \nref{sampleprofile}). 
 We see the standard gaussian random field plus a hot spot created by the coupling to a massive particle. (a)
  A slice of the distribution centered on the center of the hotspot. 
The orange line represents an instance of the fluctuating field with no hotspot. The green line shows the 
hot spot becoming larger than the mean value of the fluctuations, represented by the horizontal blue line. 
(b) Two dimensional plot of the hot spot plus the quantum fluctuations. We clearly see
the hot spot standing out over the background. This figure is zoomed relative to figure \ref{axionandmasses}.  }\label{hotspots}
\end{center}
\end{figure}

       We have postulated the existence of massive particles whose mass depends on the inflaton. This implies that there is a coupling between the inflaton and
       these massive particles. We can think of the massive particles as a classical source for the inflaton field. This produces a perturbation of the inflaton 
       around the location of the particles \cite{Itzhaki:2008ih,Fialkov:2009xm}.
  In other words, 
  the massive particles ``pull'' on the inflaton, locally delaying its evolution.   
 This in turn delays the end of inflation, causing a  further expansion in this region. 
       In order to have an observable ``classical'' signal,  we want the net effect of this pull to be larger than the quantum fluctuations. 
       In the approximation that $H$ and 
the slow roll parameter $\epsilon$ are constant, there is a surprisingly simple expression for the effect of the massive particle. 
       We find that the late time expectation value of the curvature fluctuations
 $\zeta(x)$ due to the presence a  particle at 
$\vec x =0$  is given by (see appendix \ref{HotSpot})
       \be \la{zetaclass}
        \langle \zeta_{part}(x) \rangle  =  { m(\eta = -|x|) \over 2  \sqrt{ 2 \epsilon}  M_{pl} } \times 
        \left ( { 1 \over 2 \pi \sqrt{2 \epsilon} } { H \over M_{pl} } \right) 
        \ee
        where $M_{pl}$ is the reduced Planck mass.  Notice that the mass is evaluated at a  conformal 
         time equal 
        to the distance in comoving coordinates from the location of the particle.
 Note that the time dependence of the mass is translated to 
        the spatial dependence of the profile of the field.  The last factor in \nref{zetaclass} is the amplitude of the quantum fluctuations. 
        Therefore we want the first  term to be larger than one. This can be achieved if $m\sim M_{pl}$ and $\epsilon $ is small. It would be unreasonable to 
        postulate a mass much larger than $M_{pl}$ since that would become a black hole. But if $\epsilon$ is  $10^{-3}$, then we can have a classical effect which 
        is ten times larger than the quantum fluctuations and should be visible. 
       See figure \ref{hotspots} for simulations of this classical solution plus the quantum fluctuations. 
We call this region which has a value of the primordial curvature fluctuation
 $\zeta(x)$ larger than the average a ``hot spot'', regardless of how
how it will appear in the CMB or other probe of the primordial fluctuations\footnote{
In \cite{Fialkov:2009xm} it was argued that depending on whether their size is larger than the size of
the horizon at recombination they can appears as hot or cold spots in the CMB.}.
  We want a situation where the hot spot is recognizable on an individual basis. This is the reason we required
 \nref{zetaclass}  to be larger than the quantum fluctuations.  

    \subsection{ The measurement} 
    \label{measurement} 
    
   Here we want to consider a process whose outcome depends on the isospin of the particle. Of course,
 we need to introduce an isospin breaking 
   interaction. We postulate that the mass of the massive particle has a component that depends on the isospin projection along an axis that depends on the
   axion field. 
   More precisely, we imagine that the field $h$ describing the massive particles has mass terms of the form
   \bea
      m_1^2(\phi) h^\dagger h  +\lambda_2 (\phi) h^\dagger ( \sigma_x\cos n \theta  +  \sigma_y \sin n\theta) h = \la{materm}
\cr
= m_1^2(\phi)  \left[ |h_1|^2 + |h_2|^2 \right] +  \left[  \lambda_2(\phi) e^{ i n \theta } h_1^* h_2 +  c.c. \right] 
      \eea
In the first line we view  $h=(h_1,h_2)$ as an isospin doublet bosonic field with  
the $\sigma$ matrices are acting on the isospin indices of $h$.
 In the second line we wrote the lagrangian in terms of the two
 complex component fields.\footnote{ This 
is the most general term that we can write down that is consistent with the 
$O(2)$ symmetry generated by phase rotations of the doublet, together with the ``reflection'' 
$h_1 \to h_2^*~,~~~h_2 \to h_1^* $. Also the coupling to the axion preserves a 
common symmetry under changing $h_1$ and $h_2$ by opposite phases together with a shift of the axion.} .
 The number $n$ is an integer and a value of about $n\sim 10$ is reasonable to amplify the fluctuations
shown in figure \ref{axionandmasses}.  We  introduced it 
      with the sole purpose of amplifying the effects of the fluctuations of the axion so that the 
the vector along which we are projecting the spin ranges over all possible orientations.\footnote{ We could have set $n=1$ and then produced larger fluctuations of the axion than in figure \ref{axionandmasses}, 
by decreasing the overall value of $f_a$. 
This would have the problem of 
 producing domain walls after the end of inflation. However,  we could set the post-inflationary 
      axion potential to zero and imagine that the axion is observable as a variation of the some of the fundamental constants in the late universe.   }
 The two eigenvalues of  the mass are
       \be \la{masssign}
        m_{\pm } = \sqrt{ m_1^2(\phi) \pm |\lambda_2(\phi)| } 
        \ee
       We want $   |\lambda_2 | \leq m_1^2 $ so as not to have an instability.
 We also want that at late times $|\lambda_2| $ is similar to $m_1^2$ so that the
       two eigenvalues of the mass differ by, say, a factor of two. 
In addition, we also want that $m_\pm$ are of order $M_{pl}$. 
This is  required so  that both values of the 
mass  are observable and distinguishable, as in the discussion in 
       subsection \ref{seeingmassive}. 
       In addition, we would like that $\lambda_2$ is negligible when the particles are created so that they are really created in an isosspin singlet. 
       Then, as $m_1$ rises, then $\lambda_2$ should also rise and become large. 
   
        We recognize that this subsection seems the most contrived part of the model.

        \begin{figure}
\begin{center}
\includegraphics[scale = .7]{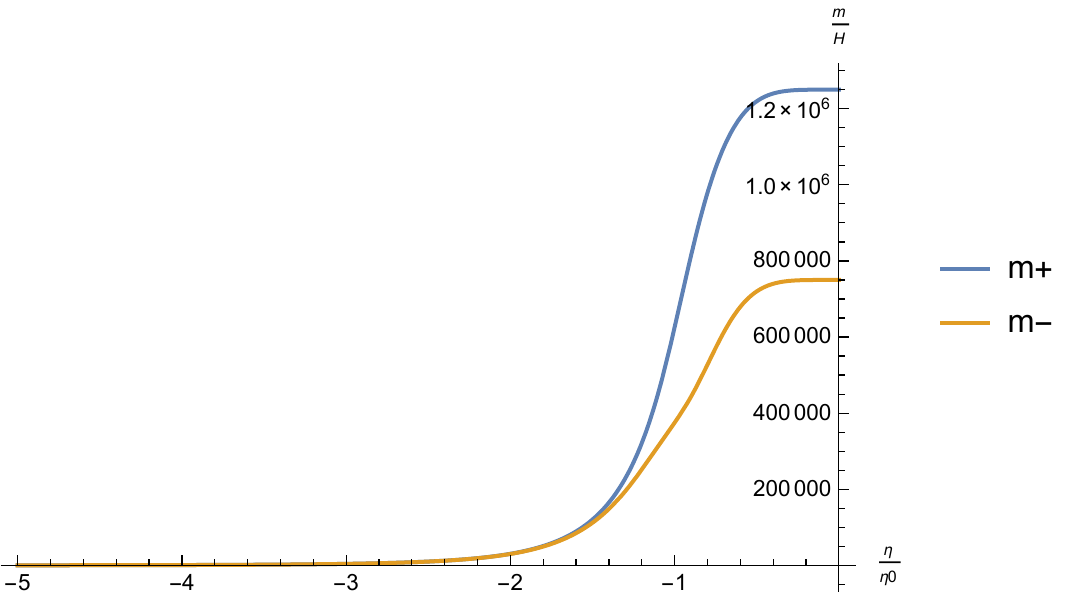} ~~~~~~\includegraphics[scale = .7]{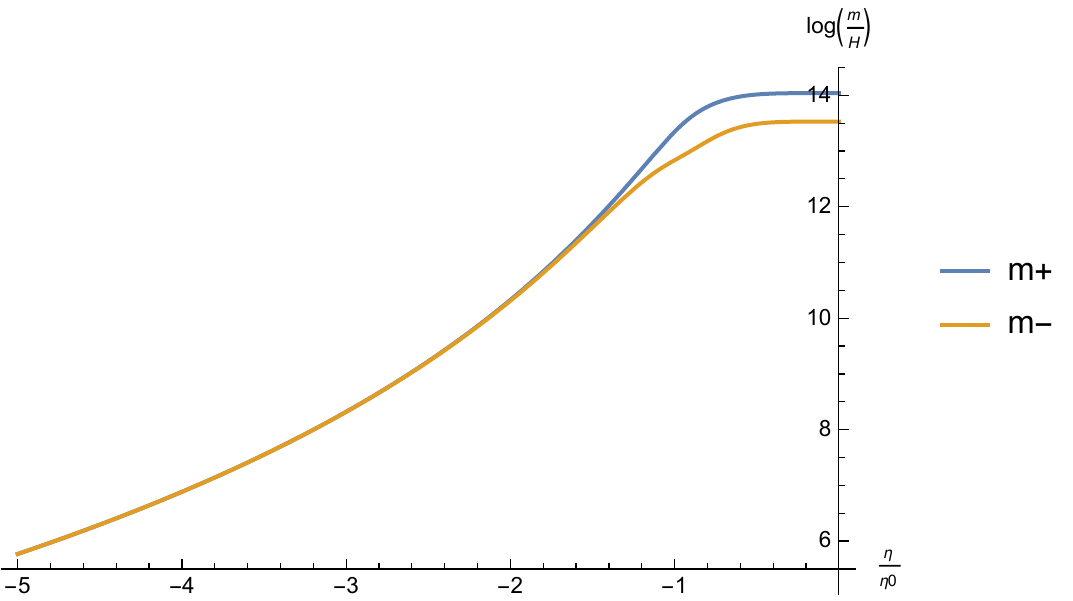}
\caption{(a) Plot of the type of $m_\pm$ functions that we want for the model.(b) Same plot in logarithmic scale. 
 We plot times after the particles are created, when  the masses are 
equal and of order $m/H \sim$ few, they then rise to large values of order $M_{pl}$ and, in addition, they become different from each other due to 
a non-zero $\lambda_2$ in \nref{materm}.   }\label{mplusmminus}
\end{center}
\end{figure}     

 Notice that in this model the field $h$ is a complex field and when we produce a particle pair one member of the pair will be a particle and
 the other an antiparticle. The complex conjugate field is, of course, also a doublet  $(h_1^*, h_2^*) = ( \tilde h_1 , - \tilde h_2)$, where the 
 $\tilde h$ field transforms as a standard doublet, in the same way as $(h_1,h_2)$ under $SU(2)$. This means, in particular, that the eigenvalues 
 of the masses are given by the projection of $ - (\sigma . \vec n)$ acting on the $\tilde h$ doublet. If we consider the pair of fields $h$ and $\tilde h$, we 
 have an ordinary spin singlet state. If $\theta $ was  constant in space, then if the particle member of the pair 
 has mass $m_+$, then the antiparticle member {\it also} has mass $m_+$. In other words, the $\pm$ of the mass of the particle is equal to the sign of the 
 $\sigma . \vec n$ operator. For the antiparticle the $\pm$ of the mass is equal to the sign of $ -(\sigma . \vec n)$ acting on $\tilde h$. 
 This extra minus sign has trivial  
 consequence, it  reverses the sign of the quantum mechanical expectation value for the $C$ observable defined in 
 \nref{Cdef} relative to what is expected for an ordinary spin singlet state.

    \subsection{ Post inflationary  observations} 
    
    We imagine that we, as  late time observers,
 can measure both the primordial scalar fluctuations as well as the primordial axion fluctuations. 
Of course, neither an axion, nor its primordial fluctuations have been seen. 
If we want to make the model consistent with present day data we can 
 imagine that the axion we are discussing corresponds to a
subleading component of the dark matter.
This dark matter density depends on the value of $\theta$ left over from the end of inflation, since
this determines the deviation from the minimum of the axion potential. Therefore, fluctuations in
$\theta$ translate into   fluctuation of this component of the dark matter. 
This is an isocurvature fluctuation. These have a characteristic scale which is set by the comoving scale 
corresponding to the time during inflation where $f_a$ was small. This is the scale of the features in 
figure \ref{axionposition}. 
In summary, by looking at the fluctuations in the subdominant matter distribution we could make a plot
of the primordial axion position at the end of inflation. The plot would look as in figure \ref{axionposition}. 

\begin{figure}
\begin{center}
\includegraphics[scale = 0.8]{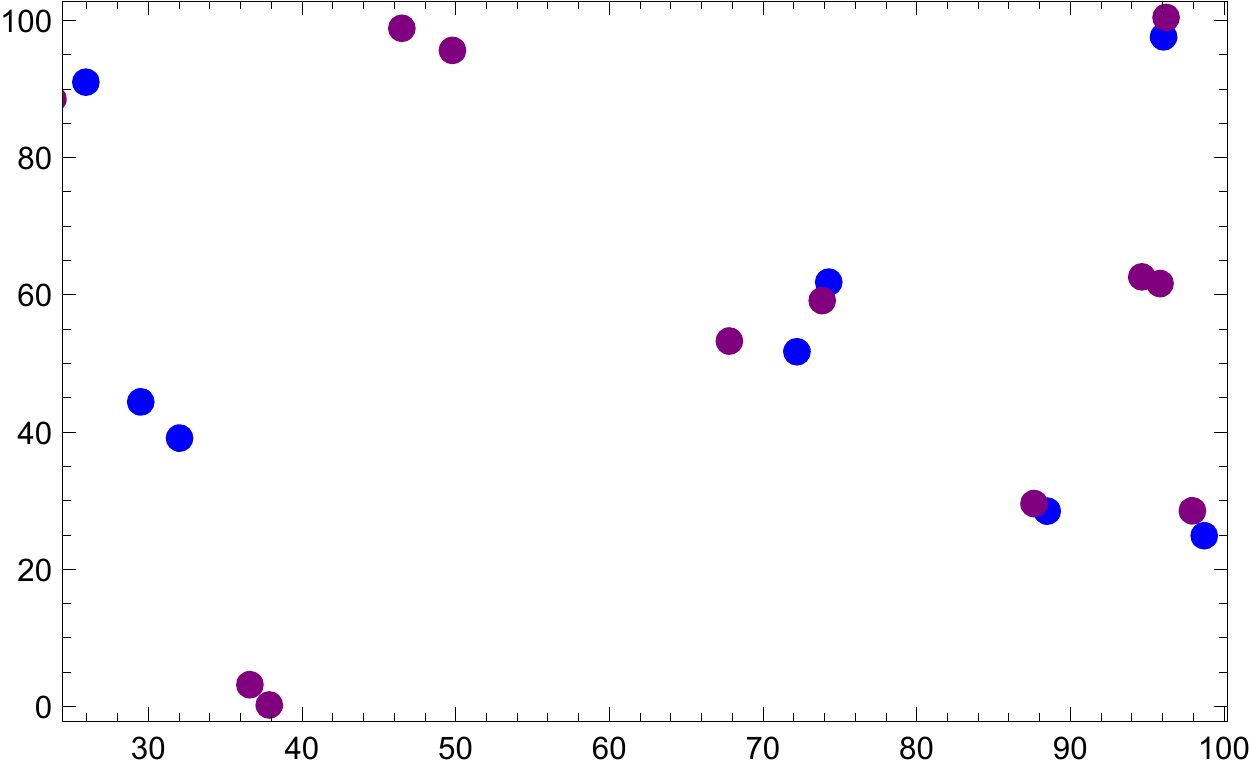}
\caption{  Performing an observation of the primordial scalar fluctuations, the observer identifies the superhot and the veryhot spots.
 Plotting only these hot spots we
get the above map. Here purple is for very hot and blue for superhot. This is the same as the map for the created particles in figure \ref{masspairs} except
that now we can associated a color (or a plus or minus sign) to each spot.    }\label{finalhotspots}
\end{center}
\end{figure}

Now let us discuss the scalar fluctuations.  In this model, 
the scalar fluctuations are given by the usual gaussian random field plus some characteristic hot spots which have a 
specific size in comoving coordinates. These hot spots are large enough to stand out from the gaussian 
field on an individual basis, as in figure \ref{hotspots}. There are two types of hot spots that differ by their 
overall amplitude. Let us call them the superhot   and the veryhot spots.
 These two possibilities correspond to the $\pm$ sign in \nref{masssign}.  
Identifying each hot spot individually and labelling it as a superhot or veryhot spot  spot we can end
up with a distribution of hot spots as  in figure \ref{finalhotspots}. 
Each hot spot can be assigned a $\pm 1$ depending on whether it is a superhot or veryhot 
spot.  Note that in this model, each hotspot corresponds to an individual massive particle created during
inflation.  The final result of this procedure is a set of pairs. And for each member of the pair we have a 
plus or minus one. We interpret this plus or minus one as the measurement of the isospin along some axis. 

Now we look  at the map of the axion angle at the location of each spot. This axion map 
could look as in figure \ref{axionandmasses}. In this way, we can assign an angle for each spot.  
    So now we have a collection of pairs of $\pm 1$ measurements together with their corresponding angles. Let us call these pairs 
    $( \pm 1_{\theta_A} , \pm 1_{\theta_B})$. These outcomes are similar to the ones obtained in the idealized Bell experiment discussed  in section \ref{ReviewBell}, where
    $\theta_{A,B}$ represents the values of the orientations of the axis along which the projection is made.  
    In other words,  we could define 
    the observable $C$ as in \nref{Cdef}, as $C(\theta_A,\theta_A' ; \theta_B ,\theta_B')$ where $\theta_A$, and $\theta_A'$ correspond to the two choices of
    detector at Alice's location and $\theta_B$, $\theta_B'$ similarly at Bob's location. We could define Alice's location to be the location of the particle and 
    Bob's location that of the antiparticle. However, in this model, the late time observer cannot distinguish between the particle and the 
    antiparticle. Fortunately, this is not a problem. In the standard Bell inequality discussion, we still have a Bell inequality if we were to consider the new observable 
    $\tilde C = { 1 \over 2} ( C + C_{ A \leftrightarrow B } )  $. And the quantum mechanical prediction for the singlet state is also invariant under $A\leftrightarrow B$, 
    since the expectation value for given choices of orientations of detectors it is proportional to  $\cos(\theta_A-\theta_B)$. 
     We can consider the angles in \nref{choices} with $\theta =\pi/4$, and looking at the observable $\tilde C$ we would observe a maximal violation of
     Bell's inequalities.\footnote{ 
      Of course, in a practical experiment we would also want to average over all configurations of \nref{choices} where the quantum mechanical 
     contributions are still the same. For example, we can perform an overall rotation or reflection of the vectors in \nref{choices}. }

  \section{Discussion} 
  
    
 Violations of the Bell inequality are   a key signature  of quantum entanglement,  
displaying the weirdness of the quantum world. 
Given that the leading theory for the origin of fluctuations in our universe relies crucially on 
quantum mechanics, it is reasonable to ask whether  
a Bell experiment is possible in cosmology. 
In tabletop experiments we have the luxury of varying the initial conditions and manipulating various 
types of materials. In cosmology, we have just the one universe we live in. However, in theory, we also 
have the ``luxury'' of imagining alternative universes where other measurements are possible. 
Here we have imagined a universe where a cosmological Bell inequality experiment is possible. 
Though the model is somewhat contrived, it shows that it is in principle possible.  
    Of course, it would be much more interesting to find such observables for the model that describes inflation in the real world (assuming that inflation does indeed
    describe the real world). 

    The exercise of constructing a model where the measurement is possible has 
 exposed some of the assumptions that are necessary in 
order to formulate an inequality. 
    In order for the observable to be subject to a Bell inequality, one needs to make several assumptions. We need to assume that the fluctuations 
    are generated when modes cross the horizon and not earlier. In other words, we need to assume that the fluctuations that generated random values of
    the axion field at the location of the two particles are not correlated with the entangled isospin state of the massive particles. This is an assumption we
    always make when we perform a Bell inequality experiment. One can question this assumption, both for today's experiments as well as for experiments 
    in the early universe, since the whole observable universe was once in a very small region of space. Nevertheless, the assumptions that go into the
    cosmological Bell inequality seem qualitatively similar to the ones that go into present day Bell inequality experiments. 
    
      Notice that a value of $|C| > 2$ implies a violation of Bell's inequalities, ruling out local classical hidden variables, 
      while a value larger than $ 2 \sqrt{2}$ would be a violation of quantum mechanics. Therefore
one can also view  it as a test of quantum mechanics.  
    
    In the model described in this paper, all the elements of the Bell experiment are constructed out of vacuum fluctuations during inflation.  
    We have seen that this can be done considering several off-the shelf elements. In fact,  particles whose mass depend on the inflaton 
    were discussed in e.g. \cite{Chung:1999ve,Kofman:2004yc}, and they arise naturally in models with moving branes as open strings stretching between the branes \cite{Kofman:2004yc}. 
Models with many fields
    are common, as well as models where an axion has quantum fluctuations during inflation, or a 
decay constant that is time dependent, see e.g \cite{Fairbairn:2014zta}.
 Perhaps the most contrived aspect was the particular 
    coupling assumed in section \ref{measurement}.

Independently of the motivation for this paper,  
 it is also interesting that particles that become very massive can leave
    a discernible signal on the spectrum of primordial fluctuations \cite{Itzhaki:2008ih,Fialkov:2009xm}.
 These are signals which 
appear on an {\it individual } basis. In other words, each hot spot is produced by 
 an individual  massive particle. 
The mass of the particle rises with the advance of the inflaton,  slowing down the inflaton around
the particle . The expansion of the universe 
imprints this signal  over  long
distances,  distances that are much larger than the Hubble radius at the end of inflation.  
The presence of particles whose mass varies so strongly with the inflaton  makes one wonder whether the potential will remain flat after the quantum 
effects of this particles are taken into account. Particles that become light at some point during the evolution are natural in monodromy inflation models 
\cite{McAllister:2008hb,Silverstein:2008sg}.  For the purposes of this paper we are content with fine tuning the potential, 
since we are not trying to argue that this particular model is the most natural one. 
 It is also  tempting to speculate that this mechanism or some  variation 
could be used to produce primordial black holes from suitably large hot spots.


Of course,  researchers 
    became convinced by quantum mechanics much before Bell inequality experiments were performed \cite{Aspect}. 
    Similarly,  there are many other features of the cosmological fluctuations that could be observed in the not so distant 
    future which would give great evidence for a quantum origin of the cosmological fluctuations. These are quantities which we compute
    using the quantum theory such as the scalar three point function \cite{Maldacena:2002vr}  (see   \cite{Chen:2010xka} for a review), 
     which could perhaps be observable using 21 cm observations \cite{Loeb:2003ya} or 
    other yet to be discovered way to measure a large number of primordial fluctuations.    
    Other possibilities include seeing oscillations in the three point function with the patterns produced by the creation of massive particles
 \cite{Chen:2009zp,Baumann:2011nk,Assassi:2012zq,Noumi:2012vr,Arkani-Hamed:2015bza}. 
 Of course, it would be interesting to find other observables that harder to reproduce using non-quantum evolution.

    Finally,   note that nature has  indeed produced a universe where Bell experiments are possible: they are certainly possible in the current
     era  of accelerated expansion, though we do not know if our results will be seen by
 any  ``post-inflationary''  observers!.

{\bf Acknowledgements } 

We would like to thank M. Zaldarriaga for an initial collaboration on the topic of this paper.  
We also thank N. Arkani Hamed, M. Mirbabayi, R. Sundrum, T. Vachaspati  and M. Simonovi{\' c} for discussions. 
I also thank D. Stanford for pointing out a problem with the plots in the first version.

J.M. is supported in part by U.S. Department of Energy grant
de-sc0009988.

 \appendix
 
 \section{Creation of massive particles with time dependent masses} 
 
 We start from the equation for a massive particle \nref{massparteq}. 
 We then define $u = h/\eta$ to find the equation 
 \be 
  u'' + p^2(\eta) u =0 ~,~~~~~~~ p^2(u) = k^2 + { m^2(\eta)/H^2  - 2 \over \eta^2 } 
 \ee
 In the standard WKB approximation the solutions are  
 \be
 u = { 1 \over \sqrt{ 2 p } } \exp\left[ i \int^\eta d\eta' p(\eta') \right] ~,~~~~~~~~ \bar u =  { 1 \over \sqrt{ 2 p } } \exp\left[ -i \int^\eta d\eta' p(\eta') \right] 
 \ee
 If the field is expanded with respect to these solutions we do not find any particle creation. 
 The WKB approximation is correct a very early and very late times. 
 The particle creation is described by finding the Bogoliubov $\beta$ coefficient which gives 
  the amount of $\bar u $ solution at late times if we start
 purely with $u$ at early times.   We will consider a situation where the WKB approximation is correct to leading order throughout the evolution. This happens when $p'/p^2 \ll 1$ at all times. 
 In this situation the particle creation will be small and it can be computed approximately using the formula 
 \be
  \beta(k) = \int_{-\infty}^0 d\eta { p'^2 \over 4 p^3 } \exp \left[ { 2 i \int_{-\infty}^\eta d\eta' p(\eta')  } \right]
  \ee
  In the situation described in  section \ref{CreationPairs} 
 we find that $\beta$ is small. But for large $k |\eta_0|$ it is even smaller because the WKB approximation is very valid for
  all times. While for $k |\eta_0| < 1$ the WKB approximation is less strongly valid near $ \eta \sim \eta_0$. But at this time we can neglect the $k$ dependence. 
  This means that we will get a $\beta(k)$ which is small and $k$ independent for $k < |\eta_0|$ and which will become even smaller for larger values of $k$. 
  A toy model for this is the function $\beta =  \epsilon e^{ - k^2 \eta_0^2 }$, with a small $\epsilon$. Here $\epsilon$ will characterize the distance between the 
  pairs of created particles while $|\eta_0|$ characterizes their relative separation. 

\subsection{An explicit example} 
\la{ExplicitMass}

   Now let us work out an explicitly solvable example. Let us assume that we have a mass that varies as 
   \be \la{toymod}
   { m^2 \over H^2 }  = \gamma  \left({ \eta\over \eta_0}  - 1\right)^2 + \delta
   \ee
   We can write down the massive wave equation \nref{massparteq} and solve for the correctly normalized solutions with definite frequency in the far past 
   \bea
   h &=& ( -\eta )^{3\over 2} x^{ - i \mu} e^{-i x} e^{ { \pi \over 2 } ( \nu + \mu) } U(\half - i \nu - i \mu, 1 - 2 i \mu ;  2 i x )  \la{hvalu}
   \cr
   {\rm with } & ~& ~ x \equiv   { \eta \over \eta_0}  \sqrt{k^2 \eta_0^2 + \gamma} ~,~~~~~\mu^2 \equiv \gamma  + \delta - { 9 \over 4 } ~,~~~~ \nu  \equiv
    {  \gamma \over \sqrt{ \gamma + k^2 \eta_0^2  } } 
   \eea
   where $U$ is the function defined as HypergeometricU in {\it mathematica}. 
    It behaves as  $U(a,b;z) \sim z^{-a}$ for large $z$. For small values of $x$,  \nref{hvalu} goes as 
   \be
  h \sim   (-\eta)^{3/2} \left[  x^{i \mu} e^{ -  \pi \mu} 2^{ 2 i \mu} {\Gamma( - 2 i \mu) \over \Gamma( \half - i \nu- i \mu) } + x^{-i\mu} { \Gamma( 2 i \mu) \over \Gamma( 
\half - i \nu + i \mu ) } \right] e^{ { \pi \over 2} (\nu + \mu) }   
\ee
For large $\mu$ and $\nu$ we get (up to irrelevant phase factors in each term) 
\be
{ 1 \over \sqrt{ 2 \mu } } e^{ - { 3 \over 2} t }  \left[ e^{ - i \mu t}e^{- \pi (\mu - \nu) }  + e^{ i \mu t} e^{- \pi/2 (\mu- \nu - |\mu-\nu| ) } 
   \right]  
\ee
with $\eta = -e^{-t}$. The first term reflects the possibility of particle creation  so that the  $\beta$ Bogoliubov coefficient is $\beta \sim e^{ - \pi (\mu - \nu) } $. 
We are interested in the case where $\mu > \nu$ so that the first term is small. Notice that the only term that depends on $k$ is in $\nu$. And $\nu$ is maximal 
for $k=0$ and it then decreases rather quickly as $k \eta_0 \sim \sqrt{\gamma}$. 
We can easily take $\gamma $ and $\delta $ to be of order one. In this way we can ensure that we produce well separated pairs. 
However, in this case, the mass does not grow enough to be visible to the late time 
observer. Therefore,  after the particles are created we need another term in the mass that makes them grow more quickly so that 
they grow to values of order  $M_{pl}$. This would require a modification of the time dependence of the mass, relative to \nref{toymod}
 for times which are a few efolds after $\eta_0$.  This new time dependence should make the mass rise to values of order $M_{pl}$ and be isospin dependent, as it is shown in figure \ref{mplusmminus}.

                    \section{ Axion with varying decay constant} 
     
\begin{figure}
\begin{center}
\includegraphics[scale = .6]{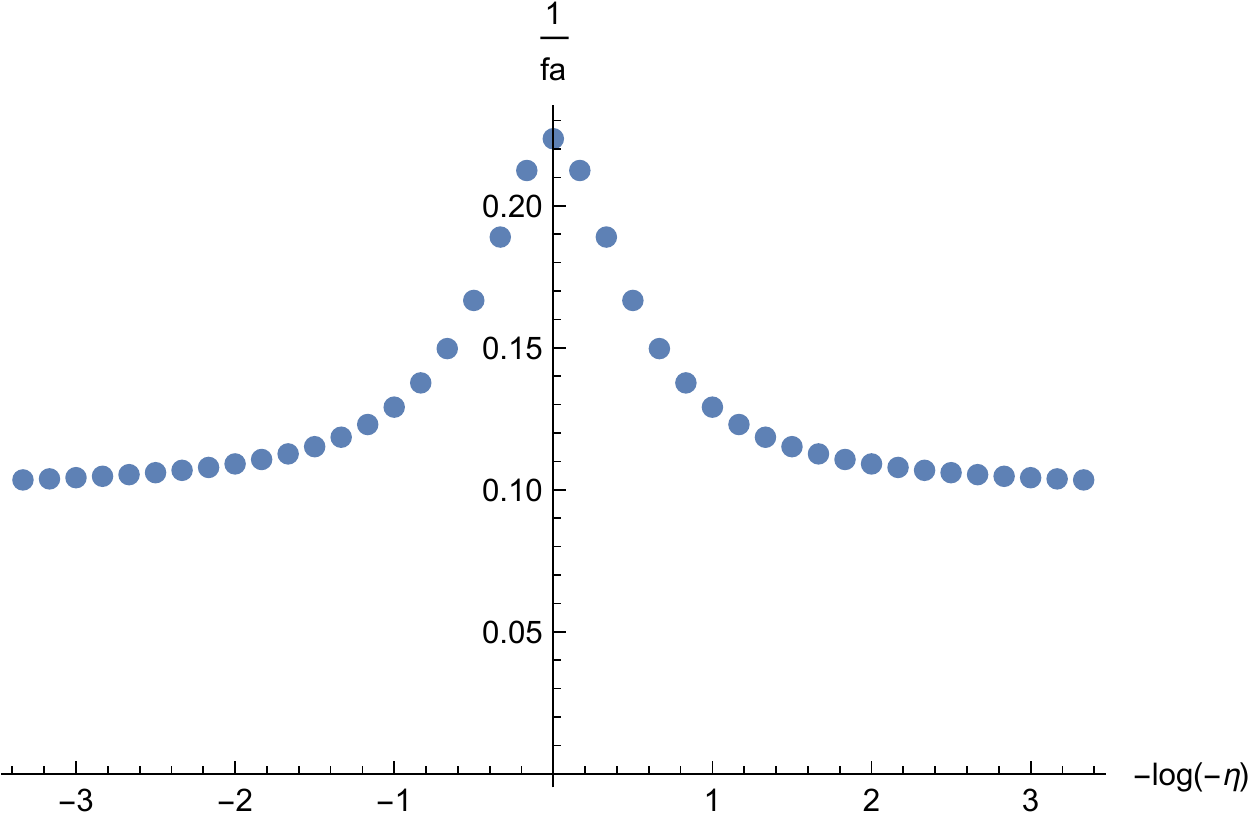}~~~~~\includegraphics[scale = .6]{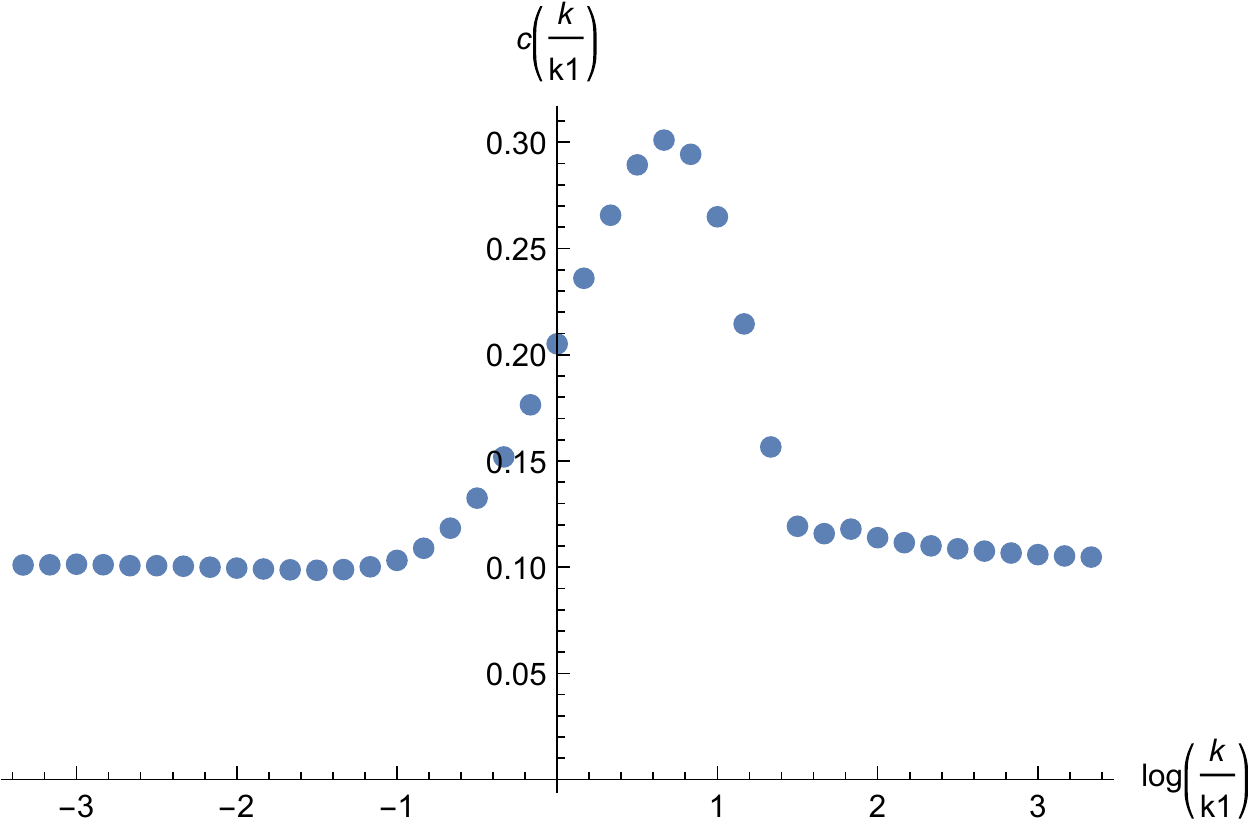}
 (a) ~~~~~~~~~~~~~~~~~~~~~~~~~~~~~~~~~~~~~~~~~~~~~~~~~(b)
\caption{ (a)  We plot of $1/\tilde f_a(\eta) $. We are plotting the inverse rather than the function itself since we expect that the axion
fluctuations scale roughly as this value. (b) Numeric computation of the axion spectrum. We plot the function $c(k)$  
defined in \nref{cdefi}.   }\label{fandcplots}
\end{center}
\end{figure}

\begin{figure}
\begin{center}
\includegraphics[scale = .9]{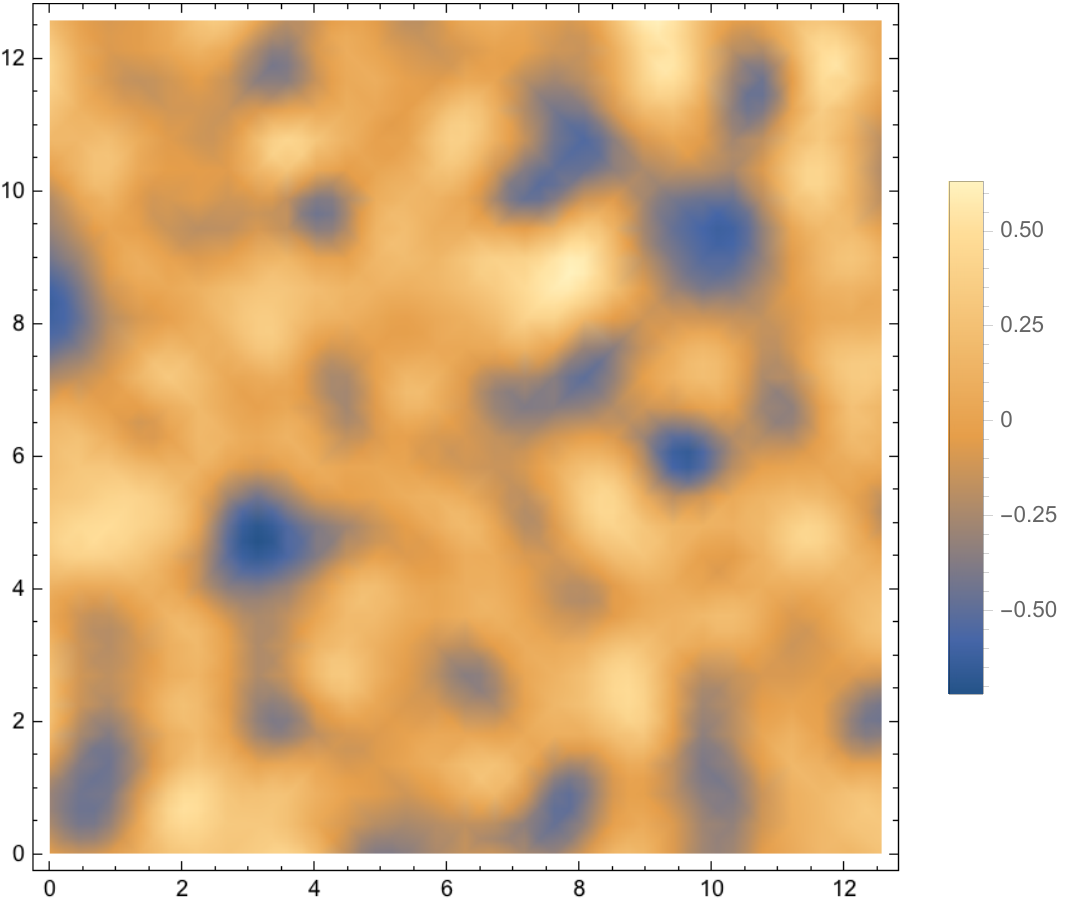} 
\caption{ We see the position space profile for the axion generated by a particular instance of
a distribution with statistics given by \nref{cdefi}, with $c(k)$ as in figure \ref{fandcplots}.  }\label{axionposition}
\end{center}
\end{figure}

     In this section we consider the fluctuations produced by an axion with varying $f_a$. 
     We consider the action \nref{actaxion}. Its equations of motion are 
     \be \la{equax}
     \partial_\eta \left[ { \tilde f_a^2 \over \eta^2 } \partial_\eta \theta \right] + { k^2 \tilde f_a^2  \over \eta^2 } \theta = 0 
     \ee
 where $\tilde f_a = f_a/H$.       As a simple example, consider the following 
\be
 \tilde f_a^2 = 100 - {  80 \over  1 + ( \log { \eta \over \eta_1 } )^2  } 
\ee
For very early or very late times this is constant and equal to $\tilde f_a \sim 10$, but 
$\tilde f_a$ dips to smaller values for time $\eta \sim \eta_1$.  We can now compute the axion fluctuations as usual. 
Namely, we  numerically solve \nref{equax} with boundary conditions at large $\eta$ corresponding to the vacuum of a harmonic oscillator and 
then look at the value of the solution for very small $\eta$. Squaring it we obtain 
the  the value of the axion fluctuations $c(k)$ defined
by 
\be \la{cdefi}
 \langle \theta(\vec k) \theta (\vec k') \rangle_{\eta \to 0}  = (2 \pi)^3 \delta^3( \vec k + \vec k') { c(k)^2 \over 2 k^3 } 
\ee
In figure \ref{fandcplots}  we see a plot of $c(k)$ for a few values of $k$. We also show in figure \ref{axionposition} the
type of position space profile for the axion generated by this probability distribution. We see that
it has features at characteristics scales but it is smooth at shorter distances.

 \section{ Effect of massive particles on the scalar curvature fluctuations } 
     \la{HotSpot}

Here we consider the Lagrangian 
\be 
S = { 1 \over 2} \int d\eta d^3 x  { 2 \epsilon M_{pl}^2 \over H^2  } \left[ { ( \partial_\eta \zeta )^2 - ( \partial_i \zeta)^2 \over \eta^2 }\right]  - \int  { d\eta \over H } m(\eta ) \partial_\eta \zeta(\eta,\vec x =0 ) 
\ee
This is the lagrangian corresponding to the curvature fluctuations. The coupling to the the massive 
particle at rest  is obtained from the coupling to $g_{00}$ which is through 
$ {  \delta g_{00}  \over g_{00}}=  \dot \zeta/ H $ (see equation (2.10) in \cite{Maldacena:2002vr}). 
We view the coupling to the mass as a perturbation, then we can apply the in-in formalism to compute the expectation value of $\zeta$ which is given 
in Fourier space by  
\bea
 \langle \zeta_{\vec k}(\eta =0)  \rangle & = & - i \int_{-\infty}^0 d\eta { m(\eta) \over H } 
  \langle \zeta_{\vec k}(0) \partial_\eta \zeta_{-\vec k}(\eta) \rangle  + c.c. 
 \cr
 & = &  { H \over 2 \epsilon  M_{pl}^2  } \int_{-\infty}^0 d\eta m(\eta) { \eta \over k } \sin k \eta 
\eea
where we assumed that $H$ and $\epsilon$ were constant for simplicity.  
Then going to position space we find that the inverse fourier transform produces a $\delta( \eta + |\vec x | )$ so that the final answer is 
\be  \la{zetaposap}
\langle \zeta(\vec x ) \rangle = { 1 \over 4 \pi } m(\eta = -|\vec x | ) \left( { H \over 2 \epsilon M_{pl}^2  }\right)
\ee
The standard gaussian  two point function in position space   is 
\be
\langle \zeta(\vec x) \zeta(0) \rangle = \left( { H^2 \over 2 \epsilon M_{pl}^2  }\right)  
\int { d^3 k \over ( 2 \pi)^3 } e^{ i \vec k . \vec x } { 1 \over 2 k^3 } = { 1 \over ( 2 \pi)^2 } \left( { H^2 \over 2 \epsilon M_{pl}^2  }\right) \log( L/|\vec x|) 
\ee
where $L$ is an IR cutoff. 
 Which naturally inspires us to write  \nref{zetaposap} as in \nref{zetaclass}.
 
 Figure \ref{hotspots} is based on choosing a  classical profile 
 \be \la{sampleprofile}
  \langle \zeta(x) \rangle = { 1 \over 2 \pi } { 5 \over (1 + ( 50 \eta)^2 } 
  \ee
   This does not quite 
 correspond to the time dependence of the mass discussed in \nref{ExplicitMass}, or figure 
\ref{mplusmminus}.  It is just an example.

\end{document}